\documentstyle[12pt,a4]{article}
\title{Unitary Gauge,\\St\"uckelberg Formalism\\ and Gauge Invariant
Models\\  for Effective Lagrangians}
\author{Carsten Grosse-Knetter\thanks{Supported in part by Deutsche
Forschungsgemeinschaft, Project No.: Ko 1062/1-2}\hspace{2mm}\thanks{E-Mail:
knetter@physf.uni-bielefeld.de}
\\and\\Reinhart K\"ogerler\\[5mm]Universit\"at
Bielefeld\\Fakult\"at f\"ur Physik\\D-4800 Bielefeld 1\\Germany}
\date{BI-TP 92/56\\December 1992}

\newcommand{\phr}[1]{Phys.\ Rev.\ {\bf #1}}
\newcommand{\phrd}[1]{Phys.\ Rev.\ {\bf D#1}}
\newcommand{\phrl}[1]{Phys.\ Rev.\ Lett.\ {\bf #1}}
\newcommand{\nphb}[1]{Nucl.\ Phys.\ {\bf B#1}}
\newcommand{\phlb}[1]{Phys.\ Lett.\ {\bf B#1}}
\newcommand{\zphc}[1]{Z.\ Phys.\ {\bf C#1}}
\newcommand{\aph}[1]{Ann.\ Phys.\ {\bf #1}}
\newcommand{\phrp}[1]{Phys.\ Rep.\ {\bf #1}}
\newcommand{\ptph}[1]{Prog.\ Theor.\ Phys.\ {\bf #1}}
\newcommand{\rx}{$\rm R_\xi$}
\newcommand{\be}{\begin{equation}}
\newcommand{\ee}{\end{equation}}
\newcommand{\bea}{\begin{eqnarray}}
\newcommand{\eea}{\end{eqnarray}}
\newcommand{\ba}{\begin{array}}
\newcommand{\ea}{\end{array}}
\newcommand{\lag}{{\cal L}}
\newcommand{\nn}{\nonumber\\}
\newcommand{\nnn}{\nonumber\\ &&}
\newcommand{\tr}{{\,\rm tr\,}}
\newcommand{\eref}[1]{(\ref{#1})}
\newcommand{\df}{{\cal D}}
\newcommand{\Det}{{\,\rm Det\,}}
\newcommand{\suu}{$\rm SU(2)\times U(1)$}
\newcommand{\uem}{$\rm U(1)_{em}$}

\begin{document}
\begin{titlepage}
\maketitle
\begin{abstract}
Within the framework of
the path-integral formalism we reinvestigate the different methods of
removing the unphysical degrees of freedom from spontanously broken gauge
theories. These are: construction
of the unitary gauge by gauge fixing; \rx -limiting procedure;
decoupling of the unphysical fields by point transformations.
In the unitary gauge there exists
an extra quartic divergent Higgs self-interaction term,
which cannot be neglected if perturbative calculations are performed
in this gauge. Using the St\"uckelberg formalism this procedure
can be reversed, i.~e., a gauge theory can be reconstructed from
its unitary gauge. We also discuss the equivalence of
effective-Lagrangian theories, containing arbitrary interactions,
to (nonlinearly realized) spontanously
broken gauge theories and we show how they can be extended to Higgs models.
\end{abstract}
\end{titlepage}


\section[]{Introduction}
The purpose of the present paper is primarily to reinvestigate the various
approaches to the  unitary gauge within {\em quantized\/}
spontanously broken gauge theories (SBGTs),
thereby putting the emphasis on the connections between the different
methods and their common basis. Although most of the described techniques are
known (at least to several groups of experts) we find it worthwile to clarify
these different approaches and, especially, to analyze the powerful method of
St\"uckelberg transformations. On the basis of this analysis we then utilize
the St\"uckelberg formalism to connect effective vector-boson Lagrangians
(containing standard or non-standard vector-boson self-interactions) with
(linearly or nonlinearly realized) SBGTs.

SBGTs contain unphysical degrees of
freedom, the pseudo-Goldstone scalars. On the classical level the unphysical
fields can be removed by means of gauge
transformations, i.~e., for given values of the pseudo-Goldstone fields at
each space-time point there exists a gauge transformation (with
gauge parameters that have to be chosen as functions
of these values)  which maps the unphysical fields identically to
zero. This gauge, which is characterized by the fact that the Lagrangian
contains only ``physical fields'' is called the unitary gauge (U-gauge).

However, this naive definition of the U-gauge
cannot be applied in quantum physics which is
best seen if one uses for quantization the framework of
Feynman's path integral (PI) \cite{feyn} and the Faddeev--Popov (FP) formalism
\cite{fapo}. On the quantum level it is the generating functional and not only
the Lagrangian which contains the complete physical
information. Since the gauge transformation which removes the unphysical fields
is dependent on the values of these fields, it cannot be applied to the
generating functional where a functional integration over {\em all\/}
values of the fields is performed. In other words there is not a ``universal''
transformation which sets arbitrary pseudo-Goldstone fields equal to zero.

There are three (equivalent) ways of constructing a gauge without unphysical
fields (i.~e.\
without pseudo-Goldstone and without ghost fields) within the
PI formalism. We discuss them mainly
for the case of linearly and minimally realized
SBGTs (i.~e.\ those which contain physical Higgs scalars \cite{higgs,kibb} and
which are renormalizable). The case of nonlinear and/or nonminimal realizations
will also be explained at the end.

The first procedure for constructing  the U-gauge
is simply to impose the gauge-fixing condition that the
pseudo-Goldstone fields are equal to zero \cite{lezj}
(which can be done because of the
existence of the abovementioned gauge transformation). The
corresponding FP $\delta$-function
is used to integrate out the unphysical scalars while the FP determinant can be
exponentiated without introducing ghost fields and yields a quartic divergent
(i.~e.\ proportional to $\delta^4(0)$) nonpolynomial Higgs-self-coupling term.

The second method is to construct the \rx -gauge
\cite{lezj,fls,able,wein2} in which unphysical fields are still
present but with masses proportional to the free parameter
$\sqrt{\xi}$, and then to perform the limit $\xi\to\infty$ \cite{wein2,leya}.
In this
limit, the unphysical fields get infinite masses and decouple. However, the
ghost-ghost-scalar couplings get infinite, too, with the consequence, that
the ghost term does not completely vanish: there remains
the abovementioned Higgs-self-coupling term.

The third way is most similar to the classical treatment: the unphysical fields
are decoupled from the physical ones by point transformations \cite{lezj}.
This procedure consists of two subsequent transformations; first the unphysical
scalars are paramatrized nonlinearly and then they are decoupled and can be
integrated out. Since in this formalism
transformations of the functional integrand are
performed, the Jacobian determinant due to the change of the integral measure
has to be considered \cite{lezj,sast},
it yields again the new Higgs-self-interaction term.

Thus, all three methods lead to a quantum level Lagrangian (in the U-gauge)
which contains, in addition to the classical U-gauge Lagrangian,
the extra nonpolynomial quartic divergent Higgs-self-interaction
term. The same term was derived by  quantizing the classical U-gauge Lagrangian
canonically \cite{leya,wein1}
where it emerges as a remnant of covariantrization.
It has been shown on the one-loop level for three- and
four-Higgs-interaction amplitudes that the quartic divergences
cancel against quartic divergent N-Higgs-vertices which are quantum induced by
gauge-boson loops in the U-gauge \cite{jogl}; this ensures renormalizability.
In fact, a linear SBGT is renormalizable even in its U-gauge because of the
equivalence of all gauges \cite{lezj} (although this is not expected by naive
power-counting due to the bad high-energy behaviour of the gauge boson
propagator in this gauge). So loop calculations can be performed
consistently in the unitary
gauge if the extra term (which does not contribute
to the presently phenomenological most interesting processes at one-loop level)
is taken into account.
Loop calculations may be simpler in the U-gauge than in the \rx -gauge
because there are less Feynman diagrams to be considered, on the other hand the
resulting expressions can be more complex because of the form of the
vector-boson
propagator which is proportional to the zeroth instead of
the inverse second power of the energy.

The third of the abovementioned procedures can be reversed, i.~e.\ a
SBGT can be ``reconstructed'' from its U-gauge Lagrangian
(which is considered as an
effective Lagrangian). To do this, scalar fields, which are initially
completely decoupled, are introduced
to the theory by multiplying an (infinite) constant to the generating
functional, which contains the
functional integration over these fields.
The unphysical scalars are then coupled to the physical fields by an
appropriate point
transformation. At the next step unphysical and physical scalar fields are
rewritten in a linearized form. This procedure
has been described in \cite{clt} and is
formulated here within the PI formalism. The method of constructing
SBGTs by such ``field enlarging transformations''
represents the non-Abelian version of the St\"uckelberg formalism
\cite{stue,kugo,sots,dtt}, which in its original form was studied only for
theories without physical Higgs bosons, where it leads
to the problem of nonpolynomial
interactions and nonrenormalizability (in non-Abelian theories). The existence
of physical scalars, however, enables a linearization of the scalar sector, so
that renormalizable St\"uckelberg models can be constructed.

One should remember in this connection
that the classical U-gauge Lagrangian can be derived simply by the
demand of tree unitarity (good high energy behaviour of tree level
cross sections) \cite{jogl,clt,llsm}:
tree unitarity implies a SBGT as
stated in \cite{clt}. However, for being able to handle quantum effects,
the Lagrangian  must contain in addition
the extra quartic divergent Higgs self-coupling term to compensate the Jacobian
determinants that arise while performing the point transformations. This term
cannot be derived from tree unitarity alone, but, as mentioned above, from
the demand of vanishing quartic loop-implied Higgs self-couplings \cite{jogl}.

Recently the St\"uckelberg formalism has attracted attention in its originial
domain of constructing Higgs-less gauge-theories, because it has been shown
that
each effective-Lagrangian theory (containing arbitrary interactions)
is equivalent to a gauge theory with (in general) nonlinearly realized symmetry
\cite{alda}; i.~e., there exists a (field enlarging) point
transformation of the fields which makes the Lagrangian gauge invariant.
For the case of massive spin-one particles with arbitrary (non-Yang--Mills)
self-interactions this equivalence has been investigated in \cite{bulo}.
We will prove that the corresponding transformation is in fact a St\"uckelberg
transformation.
Within these models the gauge group acts nonlinearly on the unphysical fields,
there are nonpolynomial interactions and they are nonrenormalizable.
However, the gauge freedom enables the choice of the \rx -gauge (where the
vector-boson propagators have a good high-energy behaviour)
to perform loop calculations in an effective-Lagrangian
theory, which shows that the loops
in such a theory do not diverge as severely as one would expect.
Besides, each
effective-Lagrangian theory with massive vector bosons
can even be extended to a SBGT with linearly
realized symmetry by introducing a physical Higgs boson. This makes the loop
corrections even smaller.

Within this paper, SBGTs are discussed by taking the example of the
$\rm SU(2)\times U(1)$ standard model (SM) of electroweak interaction
\cite{sm,sm2}
since it is of greatest phenomenological interest and since it is a
sufficiently general version
of a SBGT (the gauge group is non-Abelian, a subgroup
remains unbroken, there is gauge-boson mixing and the model contains fermions).
Special
simplifications that arise in simpler SBGTs do not take place here and results
can be easily transferred to any other SBGT.
Similarly, in our discussion of effective-Lagrangian theories
we restrict ourselves to theories containing the electroweak vector bosons,
which leads to \suu\ gauge invariance.

The paper is organized as follows: In Sect.~2 we introduce our notation of
the SM and of the FP quantization procedure.
In the next three sections we describe the
different methods of deriving the unitary gauge (within the SM):
in Sect.~3 the
removal of the unphysical scalars by gauge fixing,
in Sect.~4 the limit $\xi\to\infty$ of the \rx -gauge,
and in Sect.~5 the
decoupling of these fields by point transformations. In Sect.~6 we explain
the construction of the SM from its U-gauge Lagrangian
and reformulate the St\"uckelberg
formalism in the framework of the PI formalism. In Sect.~7 we discuss
the treatment of (nonrenormalizable) effective-Lagrangian theories
within this formalism and explain the derivation of such models from SBGTs both
with nonlinearly and with linearly realized symmetry.
Sect.~8 is devoted to a summary of our results.


\section[]{Preliminaries and Notation}
The SM gauge fields corresponding to the gauge groups SU(2) and U(1),
respectively, are $W_{i\mu}$ and $B_{\mu}$.
For practical purposes the $W$ field
is parametrized in terms of a $2\times 2$ matrix:
\be W_{\mu}=\frac{1}{2}W_{i\mu}\tau_i.\ee
The matrix valued field strength tensors are denoted by
$W_{\mu\nu}$ and $B_{\mu\nu}$.
The scalar fields $\tilde{h}$ and $\varphi_i\,(i=1,2,3)$
are written as a $2\times 2$ matrix as well:
\be\Phi=\frac{1}{\sqrt{2}}(
\tilde{h}{\bf{1}}+i\tau_i\varphi_i) .\label{higgslin}\ee
Furthermore we consider
one fermionic doublet (the generalization to more doubletts
works as usual) consisting of an up-type field $u$ and a down-type
field $d$ (quark or lepton)
\be\Psi=\left(\ba{c}u\\d\ea
\right),\qquad\Psi_{L,R}=\frac{1}{2}(1\mp\gamma_5)\Psi.\ee
The fermion mass matrix is
\be M_f=\left(\ba{cc}m_u&0\\0&m_d\ea\right).\ee
With the help of the appropriate covariant derivatives
$D_\mu\Phi$, $D_\mu \Psi_L$ and $D_\mu\Psi_R$,
the gauge invariant Lagrangian of the SM takes the well known form
\cite{sm,sm2}
\newpage
\vspace*{-9mm}
\bea\lag_{inv}&=&-\frac{1}{2}\tr(W^{\mu\nu}W_{\mu\nu})-
\frac{1}{4}(B^{\mu\nu}B_{\mu\nu})\nnn
+\frac{1}{2}\tr\left[(D^\mu\Phi)^\dagger
(D_\mu\Phi)\right]-\frac{1}{2}\mu^2\tr(\Phi\Phi^\dagger)-\frac{1}{4}\lambda\tr
(\Phi\Phi^\dagger)^2\nnn
+i(\bar{\Psi}_L\gamma_\mu D^\mu\Psi_L+\bar{\Psi}_R\gamma_\mu D^\mu\Psi_R)
-\frac{\sqrt{2}}{v}(\bar{\Psi}_L\Phi M_f\Psi_R+\bar{\Psi}_R M_f\Phi^\dagger
\Psi_L)\label{linv}\eea
(with $\mu^2<0$ and $\lambda>0$). $\lag_{inv}$ is invariant under the local
$\rm SU(2)\times U(1)$ gauge transformations
\bea W_\mu&\to &S(x)W_\mu S^\dagger(x)-\frac{i}{g} S(x)\partial_\mu
S^\dagger(x)
,\nn B_\mu&\to &B_\mu-\partial_\mu\beta(x),\nn
\Phi& \to & S(x)\Phi\exp\left(-\frac{i}{2}g'\beta(x)\tau_3\right),\nn
\Psi_L&\to & S(x)\exp\left(\frac{i}{2}g'(B-L)\beta(x)\right)\Psi_L,\nn
\Psi_R& \to & \exp\left(\frac{i}{2}g'(\tau_3+B-L)\beta (x)\right)\Psi_R,
\label{gaugetrafo}\eea
with
\be S(x)=\exp\left(\frac{i}{2}g\alpha_i(x)\tau_i\right),\ee
(where $g$ and $g'$ are the SU(2) and U(1) coupling constants and
$B$ and $L$ are the Baryon and the Lepton number of $\Psi$,
respectively). $\alpha_i(x)$ and $\beta(x)$ denote the four gauge parameters.
The nonvanishing
vacuum expectation value (VEV) of the scalar field $\Phi$ is chosen as
\be \langle \Phi\rangle_0=\frac{v}{\sqrt{2}}{\bf 1}\quad\mbox{with}\quad
v=\sqrt{\frac{-\mu^2}{\lambda}}.\label{vev}\ee
By defining \be h=\tilde{h}-v\label{higgsshift},\ee
$h$ and $\varphi_i$ have vanishing VEV. $h$ is the Higgs field and $\varphi_i$
are the pseudo-Goldstone fields.
The physical gauge-boson fields $W^\pm_\mu$, $Z_\mu$ and $A_\mu$ (photon) are
the well known combinations of $W_{i\mu}$ and $B_\mu$.

To quantize the theory one introduces the path integral \cite{feyn}
\be Z=\int\df W_{i\mu}\df B_\mu \df h\df\varphi_i \df\Psi\df\bar{\Psi}\,
\exp\left(i\int d^4x\, \lag_{inv}\right). \label{pi}\ee
It contains an infinite constant due to the gauge freedom,
which is removed in the FP formalism \cite{fapo}.
Imposing the general gauge fixing conditions
\be F_a(W_{i\mu},B_\mu,\Phi)=C_a(x),\qquad a=1,\ldots ,4\label{gf}\ee
(where $C_a(x)$ are arbitrary funcions) $Z$ is rewritten as
\be Z=\int\df W_{i\mu}\df B_\mu \df h\df\varphi_i \df\Psi\df\bar{\Psi}\,
\delta^4[F_a-C_a(x)]\Det\left(\frac{\delta F_a(x)}{\delta\alpha_b(y)}\right)
\exp\left(i\int d^4x\, \lag_{inv}\right) \label{pigf}\ee
($\alpha_a=(\alpha_i,\beta)$). Since \eref{pigf} is independent of the $C_a$
\cite{lezj}, one can perform the weighted average over them
(with the weight functions
$\exp\left(-\frac{i}{2\xi_a}\int d^4x\, C_a^2\right)$, $\xi_a$
being a set of free parameters\footnote{Usually
all $\xi_a$ are taken to be
equal, but for our purposes we allow also different $\xi_a$.\label{diffxi}})
and one expresses the FP determinant through  the
ghost fields $\eta_a^{}$, $\eta_a^\ast$ using
\be \Det\left(\frac{\delta F_a(x)}{\delta\alpha_b(y)}\right)\propto
\int\df\eta^{}_a\df\eta^\ast_a\,
\exp\left(-i\int d^4x\, \eta^\ast_a\frac{\delta F_a}{\delta
\alpha_b}\eta^{}_b\right)\label{ghosts}\ee
As a result, \eref{pigf} can be written in terms of a PI
with an effective Lagrangian \cite{lezj,fls,able,wein2}
\be Z=\int\df W_{i\mu}\df B_\mu \df h\df\varphi_i \df\Psi\df\bar{\Psi}
\df\eta^{}_a\df\eta^\ast_a\,
\exp\left(i\int d^4x\, \lag_{eff}\right), \label{pieff}\ee
where $\lag_{eff}$ is given by
\bea \lag_{eff}&=&\lag_{inv}-\frac{1}{2\xi_a}F_a^2-\eta^\ast_a\frac{\delta
F_a}{\delta\alpha_b}\eta^{}_b
\nn&\equiv&\lag_{inv}+\lag_{g.f.}+\lag_{FP}.\label{leff}\eea
Finally, source terms for all (physical and
unphysical) fields have to be added to the Lagrangian to perform perturbative
calculations\footnote{In most of the literature the source terms are added
before performing the FP procedure. We consider it as
more consistent to add them afterwards
because the source terms for the ghosts have to be added later, anyway.
However, our analysis is independent of the different ways to treat
this.\label{source}}.


\section[]{Derivation of the U-Gauge by Gauge Fixing}
In this section we explain the ``direct way'' of constructing the
U-gauge within the FP formalism by setting the unphysical pseudo-Goldstone
fields  equal to to zero from the beginning. This is done by imposing the
gauge fixing conditions \cite{lezj}
\bea \varphi_i&=&0,\nn \partial_\mu A^\mu&=&C(x)\label{ugf}.\eea
This choice is possible since the $\varphi_i$ can be transformed to zero by
gauge transformations.
The second  condition is necessary
due to the unbroken symmetry \uem , which has to be fixed as well.
\eref{pigf} now takes the form
\be Z=\int\df W_{i\mu}\df B_\mu \df h\df\varphi_i\df\Psi\df\bar{\Psi}\,
\delta^3[\varphi_i]\delta[\partial_\mu A^\mu-C(x)]
\Det\left(\frac{\delta F_a(x)}{\delta\alpha_b(y)}\right)
\exp\left(i\int d^4x\, \lag_{inv}\right) \ee
Only the second $\delta$-function is treated in the way explained above,
leading to an appropriate $\lag_{g.f.}$. The other one enables to
perform the $\df \varphi_i$ integration with the result that all
pseudo-Goldstone fields in the invariant Lagrangian  and in the ghost term
are set equal to zero. Thus, the unphysical pseudo-Goldstone
bosons are removed from the
theory. The effective Lagrangian becomes
\be \lag_{eff}=\lag_{inv}|_{\varphi_i=0}-\frac{1}{2\xi_\gamma}(\partial_\mu
A^\mu)^2+\lag_{FP}|_{\varphi_i=0}.\label{leffu}\ee
Except for $\lag_{FP}$ this is identical to the classical U-gauge Lagrangian
with fixed \uem .

Let us now derive the ghost term and show that the ghost fields can be removed
from the theory, too.
Corresponding to the gauge boson mixing we define the parameters
\bea \alpha_Z&=&\cos\theta_W \alpha_3 -\sin\theta_W \beta,\nn
\alpha_\gamma&=&\sin\theta_W \alpha_3 +\cos\theta_W \beta\eea
($\theta_W$ being the Weinberg angle defined by $\tan\theta_W=\frac{g'}{g}$).
{}From \eref{gaugetrafo} one finds the infinitesimal transformations of the
$F_a$ for the present case:
\bea \delta\varphi_1&=&\frac{g}{2}(v+h)\delta\alpha_1+O(\varphi_i),\nn
\delta\varphi_2&=&\frac{g}{2}(v+h)\delta\alpha_2+O(\varphi_i),\nn
\delta\varphi_3&=&\frac{g}{2\cos\theta_W}(v+h)\delta\alpha_Z+O(\varphi_i),\nn
\delta(\partial_\mu A^\mu)&=&-\Box\delta\alpha_\gamma+O(\varphi_i)+O(W_{i\mu})
.\label{infgauge}\eea
First, we see that all terms $O(\varphi_i)$, which are proportional to the
pseudo-Goldstone fields $\varphi_i$,
yield vanishing contributions to the ghost terms after
integrating out the $\delta$-function, as explained above.
Secondly, we note that $\eta_\gamma$ (the ghost belonging to the
electromagnetic gauge freedom) is a physically inert field: it is not possible
to construct a Feynman diagram with internal $\eta_\gamma$-lines, because
\eref{infgauge} only yields (besides a kinetic term for $\eta_\gamma$)
vertices with outgoing $\eta_\gamma$-lines (and incoming $\eta^\pm$ coupled to
$W^\pm$) but no vertices with an incoming $\eta_\gamma$. Thus, the field
$\eta_\gamma$ can be integrated out.

After removing all redundant terms, the resulting
FP determinant can be expressed as
\be \Det\left(\frac{\delta F_a(x)}
{\delta\alpha_b(y)}\right)=\Det\left(\frac{g}{2}(v+h)\, {\rm diag}
\left[1,1,\frac{1}{\cos\theta_W}\right] \delta^4(x-y)\right).
\label{detfp}\ee
Since the argument of the determinant is a local function we can
express the functional determinant (``Det'') in terms of the ordinary one
(``det'') using the relation \cite{sast}
\be \Det (M_{ab}(x)\delta(x-y))=\
\exp\left[\delta(0)\int dx\, \ln(\det M_{ab}(x))
\right].\label{det}\ee
We therefore write
\be\Det\left(\frac{\delta F_a(x)}{\delta\alpha_b(y)}\right)=\exp\left( i\int
dx\, \left(-3i\delta^4(0)
\ln\left(1+\frac{h}{v}\right)-i\delta^4(0)\ln\frac{v^3}{\cos\theta_W}\right)
\right .\ee
This means that here (in contrast to usual gauge fixing)
the locality of the FP determinant enables
exponentiation of this term without introducing unphysical ghost fields.
As a result,
we find (neglecting a constant and using $M_W=\frac{vg}{2}$)
the ghostless FP term to the Lagrangian:
\be\lag_{FP}=-3i\delta^4(0)\ln\left
(1+\frac{g}{2M_W}h\right).\label{extra}\ee
This, together with \eref{leffu}, shows that $\lag_{eff}$ (in the gauge defined
by \eref{ugf})
contains no unphysical fields neither pseudo-Goldstone nor ghost
fields\footnote{\label{foot}If the
unbroken subgroup is non-Abelian the ghost fields belonging to this
subgroup are still present.
These can be removed by choosing the axial gauge $t_\mu A_b^\mu=C_b(x)$
for the massless gauge bosons instead of the last line of \eref{ugf}.}.
Instead, there is the extra term \eref{extra} describing a quartic divergent
nonpolynomial Higgs self-interaction.

The extra term \eref{extra} can alternatively
be derived from the Feynman diagrams
obtained by expressing the determinant \eref{detfp} via
\eref{ghosts} in tems of
usual ghost fields. We are going now to present this derivation also,
since it makes the role
of the new interaction term more transparent. In this formalism the ghost term
is (introducing $\eta^\pm=\frac{1}{\sqrt{2}}(\eta_1\mp i\eta_2)$)
\bea \lag_{FP}&=&-M_W\eta^{+\ast}\eta^+ -M_W\eta^{-\ast}\eta^-
-M_Z\eta^{\ast}_Z\eta^{}_Z\nnn -\frac{g}{2}\eta^{+\ast}\eta^+h -
\frac{g}{2}\eta^{-\ast}\eta^-h -\frac{g}{2\cos\theta_W}\eta_Z^\ast
\eta^{}_Z h .\label{ufp}\eea
There are no kinetic terms of the ghost fields, but only
mass terms and couplings
to the Higgs boson. This means that the ghost propagators simply
are static ones, i.~e. inverse masses.
Figure~\ref{feynrulesug} shows the Feynman rules
derived from \eref{ufp}. Since the ghost fields only couple to the Higgs boson,
they only contribute to Feynman diagrams with internal ghost loops
connected to an arbitrary number of Higgs lines (Figure~\ref{ghostloop}),
which can be internal or external ones. Using the Feynman rules
(Fig.~\ref{feynrulesug}) the contribution of such a loop with $N$ ghost
propagators
that is coupled to $N$ Higgs bosons to the amplitude is (considering a
factor $(2\pi)^{-4}$ for the closed loop and one $(-1)$ due to the Fermi
statistics of the ghosts)
\be -\int \frac{d^4p}{(2\pi)^4}\left(-\frac{g}{2M_W}\right)^N=-\delta^4(0)
\left(-\frac{g}{2M_W}\right)^N \label{higgsloopcont} \ee
for an internal $\eta^\pm$ as well as for an internal $\eta_Z$. We see that
such a ghost loop effectively provides for a quartic divergent
$N$-Higgs
self-coupling. Let us for a moment go to the one-loop level (where
the Higgs lines connected to the ghost loop are ``tree lines'') and consider
all subdiagrams of type of Fig.~\ref{ghostloop} with a fixed number $N$  of
Higgs lines. The sum of their contributions is
\be -3\delta^4(0)(N-1)!\left(-\frac{g}{2M_W}\right)^N \label{vertexfactor}\ee
since there are three types of internal ghosts and (as one can easily verify by
induction) $(N-1)!$ different possibilities to connect the $N$ Higgs lines
to such a loop. So all the ghost loops with $N$ Higgs lines
together can be replaced by an
extra $N$-Higgs vertex (Figure~\ref{extravertex}) with the quartic divergent
vertex factor \eref{vertexfactor}. Considering a combinatorical factor of
$1/N!$ due to the $N!$ different possibilities to connect $N$ Higgs lines to
the such a vertex, all the extra Higgs vertices (with all possible  values of
$N$) can be derived from a Lagrangian
\be \lag_{extra}=3i\delta^4(0)\sum_{N=1}^\infty\left(\frac{1}{N}
\left(-\frac{g}{2M_W}\right)^Nh^N\right)=-3i\delta^4(0)\ln\left(
1+\frac{g}{2M_W}h\right),\label{extraalt}\ee
which is identical to \eref{extra}.

This one loop derivation can easily be
generalized to arbitrary loop order without changing the result:
assuming that the Higgs lines  in
Figs.~\ref{ghostloop} and \ref{extravertex} are not ``tree lines'' but
they are connected to loops among themselves or to
other ghost loops or extra vertices of this type,
the only thing in the above discussion that changes is the
combinatorics. However, the combinatorical factors of $(N-1)!$ for the
$N$-Higgs ghost loop and
of $N!$ for the $N$-Higgs vertex change by the same extra
factor so that this cancels.
One easily can see that, no matter how the Higgs legs are connected, for each
way to  attach $N$ Higgs lines to a loop like Fig.~\ref{ghostloop}
there are $N$ ways to attach them to a vertex like Fig.~\ref{extravertex},
corresponding to the $N$ cyclic permutations.

This alternative derivation of the extra term \eref{extra} to the Lagrangian
(although it is more elaborate) shows explicitly the meaning of \eref{extra}:
$\delta^4(0)$ has to be interpreted as a quartic divergent integral stemming
from
\eref{higgsloopcont} and can be expressed in terms of a cut-off $\Lambda$ as
$\frac{\Lambda^4}{(2\pi)^4}$, the logarithm has to be evaluated in a power
series
as in \eref{extraalt} and represents a (nonpolynomial) self-interaction of an
arbitrary number of Higgs bosons.
We see that the unphysical ghost fields can be effectively removed from the
theory by taking the U-gauge, but they do not completely decouple (as the
pseudo-Goldstone fields do): there remains the additional
interaction term \eref{extra} as a
remnant. However there are no more explicit ghost fields in this term.


\section[]{\rx -Limiting Procedure}
The second approach to the unitary gauge is to start from the SM in the
\rx -gauge and then to perform the limit $\xi\to\infty$.
To construct this limit, one has to modify the general
\rx -gauge a bit because the photon propagator
\be  i\frac{-g^{\mu\nu}+(1-\xi)\frac{\textstyle p^\mu p^\nu}{\textstyle
p^2}}{p^2} \label{gammaprop} \ee
would become infinite in this limit. We impose the usual gauge fixing
conditions \cite{lezj,fls,able,wein2} but express them in terms of the mass
eigenstates $A^\mu$ and $Z^\mu$ of the neutral sector
instead of $W_3^\mu$ and $B^\mu$:
\newpage
\vspace*{-9mm}
\bea  \partial_\mu
W^\mu_{1,2}-\xi M_W\varphi_{1,2}&=&C_{1,2}(x),\nn
\partial_\mu Z^\mu -\xi M_Z \varphi_3&=&C_3(x),\nn
\partial_\mu A^\mu& =&C_4(x).\label{rgf}\eea
In order to obtain the corresponding $\lag_{g.f.}$ we make use of the
possibility to introduce different parameters $\xi_a$
in $\lag_{g.f.}$ for each $F_a$ ($a=1,\ldots ,4$)
(see footnote~\ref{diffxi}).
The gauge fixing Lagrangian most convenient for our purposes is
\be \lag_{g.f.}=-\frac{1}{2\xi}\left[\sum_{i=1}^2(\partial_\mu W^\mu_i-\xi M_W
\varphi_i)^2+(\partial_\mu Z^\mu -\xi M_Z \varphi_3)^2\right]-
\frac{1}{2\xi_\gamma}(\partial_\mu A^\mu)^2\label{lgfr}\ee
with {\em two\/} free parameters $\xi$ and $\xi_\gamma$. As the ghost term
depends only on $\xi$ since \eref{rgf} depends only on $\xi$, the sole
difference of our gauge to the usual \rx -gauge is that $\xi$ in the photon
propagator is replaced by $\xi_\gamma$. Now we can take the limit
$\xi\to\infty$ \cite{wein2,leya}
(which does not affect physical observables since these do not depend on $\xi$
\cite{lezj,able})
while the unbroken subgroup \uem\ is fixed in an arbitrary
gauge specified by a finite $\xi_\gamma$
so that the photon propagator \eref{gammaprop} remains finite.

A complete list of Feynman rules for the \rx -gauged SM is, e.\ g., given in
\cite{balo}. The $\xi$ dependent parts are:
\begin{itemize}
\item The propagators of the massive gauge bosons
\be  i\frac{-g^{\mu\nu}+(1-\xi)\frac{\textstyle
p^\mu p^\nu}{\textstyle p^2-\xi M_B^2}}{p^2-M_B^2}
\ee ($M_B=M_W$ or $M_Z$), which become Proca propagators of
massive spin-one particles for infinite $\xi$.
\item The propagators of the pseudo-Goldstone bosons and the ghost fields
(except for $\eta_\gamma$, which is massless)
\be i\frac{1}{p^2-\xi M_B^2}.\ee
For $\xi\to\infty$, these particles acquire infinite mass and their propagators
vanish. If there would be no $\xi$ dependent couplings these particles
would completely decouple.
\item All couplings of a (physical or unphysical) scalar to a ghost pair.
These are proportional to $\xi$ and become
infinite in the \rx -limiting procedure.
\end{itemize}
We now classify all Feynman diagrams containing lines corresponding to
unphysical fields that do not vanish for $\xi\to\infty$.
Since the pseudo-Goldstone and the $\eta^\pm,\eta_Z$ propagators behave as
$\xi^{-1}$ for large $\xi$ their number has to equal the number of
scalar--ghost--ghost vertices ($\propto\xi$) in such diagrams. This means:
\begin{itemize}
\item All propagators $\propto\xi^{-1}$ have to be coupled to $\xi$-dependent
vertices at {\em both\/} ends, i.~e., couplings of unphysical fields to
gauge fields do not contribute in this limit.
\item Only those $\xi$-dependent vertices yield nonvanishing contributions
which couple to two $\xi$-dependent and one $\xi$-independent propagator.
These are the $h\eta\eta^\ast$ and the
$\varphi^\pm\eta^{\pm\ast}\eta_\gamma$ vertices. However, the latter do not
contribute, since they exist only with an incoming $\eta_\gamma$ but not with
an outgoing one, so it is not possible to construct closed ghost loops with
them.
\end{itemize}
As a consequence, all graphs with pseudo-Goldstone lines vanish for
$\xi\to\infty$. Therefore, the $\varphi_i$-fields can be neglected altogether
in this limit. Furthermore the only nonvanishing (sub)diagrams involving
unphysical particles are those with
ghost loops that are exclusively coupled to Higgs bosons
(Fig.~\ref{ghostloop}). The corresponding Feynman rules are given in
Fig.~\ref{feynrx}.

The contribution of such a loop with $N$ external Higgs lines
for $\xi\to\infty$ is (for internal $\eta^\pm$ as for $\eta_Z$)
\bea && -\lim_{\xi\to\infty}\int\frac{d^4p}{(2\pi)^4}
\left(-\frac{i}{2}\xi gM_W\right)^N\prod_{i=1}^N\frac{i}{p_i^2-\xi M_W^2}\nnn
=-\int \frac{d^4p}{(2\pi)^4}\left(-\frac{g}{2M_W}\right)^N=-\delta^4(0)
\left(-\frac{g}{2M_W}\right)^N.\label{looplim}\eea
(The momenta $p_i$ of the internal ghosts do not have to be specified to
perform the limit.) This is identical to \eref{higgsloopcont} and
we can transfer the
discussion of the previous section and find for the limit of the ghost term
\be\lim_{\xi\to\infty}\lag_{FP}=-3i\delta^4(0)\ln\left
(1+\frac{g}{2M_W}h\right),\ee which is exactly the extra term \eref{extra}.

Thus in the \rx -gauge, after taking the limit $\xi\to\infty$,
the pseudo-Goldstone fields and even the massless ghost field
$\eta_\gamma$, which has a $\xi$-independent
propagator, decouple completely\footnote{Remember footnote~\ref{foot}.},
while the contributions of the massive ghosts can be summarized in terms of the
extra interaction Lagrangian \eref{extra}. Therefore we obtain the same result
for the U-gauge Lagrangian as in the previous
section\footnote{In fact it is a priori
not clear that the limit $\xi\to\infty$ can be performed {\em before\/}
the loop
integration in \eref{looplim}. The fact that the obtained result is identical
to that of the alternative derivations justifies  this treatment.}.


\section[]{Decoupling the Unphysical Scalars}
In this section we derive the U-gauge Lagrangian by applying appropriate point
transformations to the invariant Lagrangian \eref{linv}. We start with
reparametrizing (point transforming)
the scalar sector of the theory \eref{higgslin} nonlinearly
\cite{higgs,kibb,lezj}:
\be \Phi=\frac{1}{\sqrt{2}}((v+h){\bf 1}+i\tau_i\varphi_i)=
\frac{1}{\sqrt{2}}(v+\rho)
\exp\left(i\frac{\zeta_i\tau_i}{v}\right).\label{nonlin}\ee
Here, $\rho$ is the new Higgs field and the $\zeta_i$ are the new
pseudo-Goldstone fields. We see that in this parametrization the Lagrangian
\eref{linv} contains nonpolynomial interactions of the $\zeta_i$ to the gauge
bosons and to the fermions, which stem from expanding the exponential
in the kinetic term of $\Phi$ and in the Yukawa
term. At the quantum level
this is not the whole story: it is not the Lagrangian
\eref{linv} but the PI \eref{pi} which is the basis of quantization.
Therefore we have to
transform the integration measure, too, which yields a functional Jacobian
determinant \cite{lezj,sast} according to
\be \df h \df\varphi_i =\df\rho\df\zeta_i\,
\Det\frac{\delta(h,\varphi_i)}{\delta(\rho ,\zeta_i)}.\ee
The explicit form of the point transformation \eref{nonlin}
is (introducing $\zeta=\sqrt{
\zeta_1^2+\zeta_2^2+\zeta_3^2}$, $\hat{\zeta}_i=\zeta_i/\zeta$
and $\tilde{\zeta}=\zeta/v$)
\bea h&=&(v+\rho)\cos\tilde{\zeta}-v,\nn
\varphi_i&=&(v+\rho)\hat{\zeta_i}\sin\tilde{\zeta}.\label{nonlintrans}\eea
Since the Jacobian matrix is a local function, we can again use \eref{det}
to express the functional determinant in terms of the ordinary one,
wich is given by
\be\det\frac{\partial(h,\varphi_i)}{\partial(\rho ,\zeta_i)}=
(v+\rho)^3\frac{\sin^2\tilde{\zeta}}{v\zeta^2}. \ee
Exponentiating this, using \eref{det},
we obtain the following extra terms to the Lagrangian due to the
change of the functional measure (after dropping a constant)
\be\lag^\prime=-3i\delta^4(0)\ln\left(1+\frac{g}{2M_W}\rho\right)
-i\delta^4(0)\ln\left(\frac{\sin^2\tilde{\zeta}}{\tilde{\zeta}^2}\right).
\label{twoextra}\ee
The first term is just our well known quartic divergent nonpolynomial Higgs
self-coupling term \eref{extra},
but there is also a quartic divergent nonpolynomial
self-interaction term of the nonlinearly realized pseudo-Goldstone fields.
The latter is not really important for practical purposes since the $\zeta_i$
become decoupled in the next step
to be performed below. (It has to be considered, however, if one
actually wants to perform perturbative calculations with this
parametrization of the scalars.)

Introducing  the field combination
\be U\equiv\exp\left(i\frac{\zeta_i\tau_i}{v}\right), \label{u}\ee
one can deduce from \eref{gaugetrafo} the behaviour of $\rho$ and $\zeta_i$
under \suu\ gauge transformations:
\bea \rho&\to&\rho,\nn
U&\to&S(x)U\exp(-\frac{i}{2}g'\beta(x)\tau_3),\label{zetatrafo}\eea
i.~e.,
the physical scalar $\rho$ is a singlet.
Consequently, the first term in \eref{twoextra} is
gauge invariant while the second is not. This is, however, no serious problem
since gauge invariance is not really
destroyed, it is just not completely obvious due
to the nonlinear parametrization\footnote{This can easily be visualized with
the help of an
example from quantum mechanics: if one studies a translational invariant
Lagrangian and transforms the functional intergration measure to polar
coordinates one finds a non-translational-invariant
extra term to the Lagrangian,
although physics is still translational invariant.}.

In order now to remove the unphysical scalars $\zeta_i$ from the theory and
thus obtaining the U-gauge Lagrangian, one can proceed in several ways. One
possibility is to apply either methods of the previous sections. We do not
perform this in dedail but only mention the main features.
If one imposes the gauge fixing condition \eref{ugf} (U-gauge) or \eref{rgf}
(\rx -gauge) with $\varphi_i$ replaced by $\zeta_i$ one finds the analogous
expressions for the ghost terms, except that there are additional interactions
of more than one pseudo-Goldstone boson with the ghost fields (which do not
affect the discussion) and that there are no couplings of the Higgs boson to
the ghosts because the transformations of the $\zeta_i$ \eref{zetatrafo} do not
depend on $\rho$.
So no ghost loops as in Figure~\ref{ghostloop} can be
constructed. Therefore, remembering our previous discussion, we see
that the ghost terms vanish
completely after integrating out $\int\df\zeta_i\, \delta^3(\zeta_i)$ in the
first case and after taking the limit $\xi\to\infty$ in the second case,
respectively.

Here, we choose one further possibility and apply
one further point transformation, which
affects the gauge and the fermion fields. It is just a reversed
(non-Abelian) St\"uckelberg transformation \cite{lezj,clt,kugo,dtt}:
\bea w_\mu&=&U^\dagger W_\mu U -\frac{i}{g}U^\dagger\partial_\mu U,\nn
b_\mu&=&B_\mu,\nn
\psi_L&=&U^\dagger\Psi_L,\nn
\psi_R&=&\Psi_R .\label{stue}\eea
The Jacobian of this transformation is  independent of the physical fields,
since \eref{stue} is linear in them. Using \eref{det} it yields the
(non-gauge-invariant) extra term
\be \lag^{\prime\prime}=-4i\delta^4(0)\ln(\sin^6\tilde{\zeta}+
\cos^6\tilde{\zeta}).\label{sinsix}\ee

We now show that the unphysical fields $\zeta_i$ decouple from the physical
fields $w_{i\mu},b_\mu,\rho$ and $\psi$. The Lagrangian \eref{linv} originally
contains couplings of pseudo-Goldstone bosons to the
gauge bosons (in the kinetic term of $\Phi$) and to the fermions (in the Yukawa
term). Expressing these terms by means of the new fields \eref{nonlin},
\eref{stue} we find
\bea&& \frac{1}{2}\tr\left[(D^\mu\Phi)^\dagger(D_\mu
\Phi)\right]-\frac{\sqrt{2}}{v}
(\bar{\Psi}_L\Phi M_f\Psi_R+\bar{\Psi}_RM_f\Phi^\dagger\Psi_L)\nn
&=&\frac{1}{2}(\partial^\mu\rho)(\partial_\mu\rho)+\frac{1}{4}(v+\rho)^2
\tr\left[\left(gw_\mu-\frac{1}{2}g'b_\mu\tau_3\right)
\left(gw^\mu-\frac{1}{2}g'b^\mu\tau_3\right)\right]\nn
&&-\frac{v+\rho}{v}(\bar{\psi}_LM_f\psi_R+\bar{\psi}_RM_f\psi_L),\eea
We see that the pseudo-Goldstone fields $\zeta_i$ have effectively disappeared
here. They only emerge in the extra terms \eref{twoextra} and \eref{sinsix},
i.~e.\ without being coupled to any other fields. We therefore can integrate
them out in the PI which yields a constant factor
\be\int\df\,\zeta_i\exp\left(i\int d^4x\,\tilde{\lag}\right)
\label{fieldenlarge}\ee
with\be\tilde{\lag}=-i\delta^4(0)\left(\ln\left(
\frac{\sin^2\tilde{\zeta}}{\tilde{\zeta}^2}\right)+
4\ln(\sin^6\tilde{\zeta}+\cos^6\tilde{\zeta})
\right).\ee
This can be removed by multiplying the PI with the compensating factor.

The effect of the point transformations \eref{stue} on the other terms in
\eref{linv} is just to replace the fields ($W_{i\mu},B_\mu,\Psi$) by
($w_{i\mu},
b_\mu,\psi$) due to the gauge invariance of these terms since \eref{stue}
formally acts on the {\em physical\/} fields as a SU(2) gauge transformation.

In summary we see that the unphysical fields $\zeta_i$ have decoupled and we
obtain the same result for the U-gauge Lagrangian as in the previous two
sections; in particular the extra term \eref{extra} is again recovered.

Next we study the behaviour of the new fields under gauge transformations.
{}From \eref{gaugetrafo}, \eref{zetatrafo}, and \eref{stue} we find that all
physical fields are invariant under the action of SU(2) and transform under
U(1) as
\bea w_\mu&\to&\exp\left(\frac{i}{2}g'\beta(x)\tau_3\right) w_\mu
\exp\left(-\frac{i}{2}g'\beta(x)\tau_3\right)-\frac{1}{2}\frac{g'}{g}
\partial_\mu \beta(x)\tau_3,\nn
b_\mu&\to&b_\mu-\partial_\mu\beta(x),\nn
\rho&\to&\rho,\nn
\psi_{L,R}&\to&\exp\left(\frac{i}{2}g'(\tau_3+B-L)\beta(x)\right)\psi_{L,R}.\eea
Introducing the mass and charge eigenstates (which are now also assigned by
small letters) $w^\pm_\mu$, $z_\mu$ and $a_\mu$
and rescaling the gauge parameter as
\be g'\beta(x)=e\kappa(x)\label{rescale}\ee one finds
\bea w_\mu^\pm&\to&\exp(\pm ie\kappa(x))w_\mu^\pm,\nn
z_\mu&\to&z_\mu,\nn a_\mu&\to& a_\mu-\partial_\mu \kappa(x),\nn
\rho&\to&\rho,\nn \psi&\to& \exp(ieQ_f\kappa(x))\psi,\label{emgauge}\eea
where $Q_f$ is the fermion charge matrix
\be Q_f=\left(\ba{cc} q_u&0\\0&q_d\ea\right). \ee
This is just an electromagnetic gauge transformation\footnote{On the first
look this seems surprising since we  performed a $\rm U(1)_Y$
transformation to derive \eref{emgauge}. However, due to the point
transformation \eref{stue}, this acts differently on the transformed fields
than on the original ones; so it becomes a $\rm U(1)_{em}$ transformation.}.
Thus, after having parametrized
the fields such that all pseudo-Goldstone bosons decouple,
the action of the whole gauge group on the physical fields reduces to
a gauge transformation belonging to the unbroken subgroup. The remaining
gauge freedom is only connected to the unphysical scalars and
has been ``removed'' by dropping \eref{fieldenlarge}.
Finally, the $\rm
U(1)_{em}$ gauge freedom has to be fixed by adding the term
\be \lag_{g.f.}=-\frac{1}{2\xi_\gamma}\partial_\mu a^\mu ,\label{gfuem}\ee
while the ghost belonging to it decouples\footnote{Remember
footnote~\ref{foot}.}, and we finally find the same effective Lagrangian
as in the previous two sections.

This derivation of the U-gauge is similar to the classical treatment  of
removing the pseudo-Goldstone scalars by a
means of a gauge transformation. In fact the
St\"uckelberg transformation \eref{stue} {\em formally}
acts as a SU(2) gauge transformation on the gauge
and fermion fields where the gauge {\em parameters}
$\alpha_i$ are replaced by the
the pseudo-Goldstone {\em fields} $-\zeta_i/M_W$.
But there remains a principal difference between that gauge transformation and
a St\"uckelberg transformation. Namely, the
gauge transformation which transforms $\zeta_i$ identically
to zero depends on the
numerical values of these fields at the various space-time points and it is not
the same transformation for different functions $\zeta_i$.
The St\"uckelberg transformation \eref{stue}, however, decouples the
pseudo-Goldstone fields (independently of their functional forms)
from the physical fields and so it can be
applied to the PI, where an integration over
all functions $\zeta_i$ is performed. The only
point where quantum physics enters the above discussion\footnote{If one
considers the source terms from the beginning (see footnote~\ref{source}),
the external sources become, after performing the point transformations,
coupled to functions of the fields instead of the fields themselves. However
this does not affect physical matrix elements \cite{able}.}
is the need of point transforming
the functional integration measure,
too. The corresponding Jacobian determinants give rise to the extra term
\eref{extra}.


\section[]{Construction of a Spontanously Broken Gauge Theory Using the
St\"uckelberg Formalism}
The formalism of the last section also can be reversed: one can start from the
U-gauge Lagrangian and construct the invariant Lagrangian by subsequent point
transformations. Although it is clear from the previous section what do to, we
explain this procedure a bit more in detail, since it is an alternative
derivation of a SBGT, which illustrates some features of such a model
quite well. Furthermore, the U-gauge Lagrangian of a SBGT can be
motivated on a rather intuitive basis: it only involves ``physical'' fields
(i.~e.\ all the fields corresponding to observable particles) and it is the
most general (effective) Lagrangian with massive vector bosons and scalars
which guarantees tree-unitarity (i.~e.\ $N$-particle $S$-matrix elements
calculated on tree level decrease at least as $E^{4-N}$ for high energy $E$)
\cite{jogl,clt,llsm}. Therefore, it seems interesting to look wether and how
the general (gauge invariant) structure of a SBGT can be reconstructed from its
U-gauge Lagrangian. Tree unitarity implies, in particular, the need of
physical scalars (Higgs bosons)
with appropriate couplings to the other particles and to itself
in order to ensure good
high energy behaviour.  It is clear that the extra term \eref{extra}, which
genuinely reflects quantum effects, is not obtained from tree level arguments,
but it is inferred from the requirement of vanishing quartic divergent
$N$-Higgs self-couplings (which has been shown for $N=3,4$ on 1-loop level in
\cite{jogl}). In the following we take the term \eref{extra} as a given part of
the U-gauge Lagrangian.

Starting from this full
U-gauge Lagrangian (containing the fields $w_{i\mu},b_\mu,
\rho$ and $\psi$ as in the previous section)
one recognizes first the local
\uem\ symmetry \eref{emgauge}. Reversing the FP procedure for the unbroken
subgroup by reintroducing the
integration over the gauge parameter as an (infinite) constant to the PI,
one can remove the g.~f.\ term \eref{gfuem} (and the ghost term if the
unbroken subgroup is non-Abelian). Next, one can see that the Lagrangian is
gauge invariant not only under \uem\ but also under the larger group \suu\
except
for the mass terms of the vector bosons and the fermions and the couplings of
the physical scalar to these particles; the kinetic terms, the
vector-boson--fermion interaction and vector-boson self-interaction terms
represent exactly an
unbroken gauge theory. The purpose of the St\"uckelberg formalism is now
to introduce, by field enlarging transformations, unphysical degrees of freedom
with an appropriate behaviour under \suu\ gauge transformations that
compensate the effect of the transformations of the physical fields in
the non-gauge-invariant terms in the U-gauge
Lagrangian \cite{clt,stue,kugo,sots,dtt}.

In the PI formalism, the field enlarging transformation is constructed
as follws: First one introduces the completely decoupled fields
$\zeta_i$ by formally multiplying
the (infinite) constant \eref{fieldenlarge} to the PI. This contains the
funtional integration over the unphysical fields $\zeta_i$ (St\"uckelberg
scalars) and an exponential, which is needed
to remove the Jacobian determinant of the following point transformations.
Then the $\zeta_i$, parametrized in terms of the unitary matrix $U$
\eref{u}, are coupled to the physical
fields by the transformations \eref{stue}\footnote{In \cite{sots} one can find
an alternative construction of the SM using the St\"uckelberg formalism.
There the point transformations do not look like SU(2) transformations,
as in our case, but like full \suu\ transformations with  gauge parameters
replaced by unphysical scalars, thus introducing four unphysical fields.
However, there exists a reparametrization of these scalars (i.~e.\ one more
point transformation) that decouples one of these fields, so that one finally
obtains the same result as we do.}.
Defining the behaviour of $U$ under \suu\ as in \eref{zetatrafo}, one finds
that the transformations of $W_{i\mu},B_\mu$ and $\Psi$, which arise from the
original \uem\ gauge freedom of the $w_{i\mu},b_\mu$ and $\psi$ (\eref{emgauge}
with \eref{rescale}) and from the transformations of $U$, are exactly the usual
gauge transformations \eref{gaugetrafo}. The Lagrangian is (except for the
extra term contained in \eref{fieldenlarge}), gauge invariant under these
transformations, because the fields appear only in the combinations
\eref{stue}. The effect of an arbitrary gauge transformation
(\eref{gaugetrafo},
\eref{zetatrafo}) on these is just an electromagnetic one \eref{emgauge}.
Since the St\"uckelberg transformation
\eref{stue} has the form of a SU(2) gauge
transformation,  its effect on the kinetic terms, vector-boson self-interaction
term and vector-boson--fermion interaction terms is just to replace $W_{i\mu}
\to w_{i\mu}$, etc. (because
these terms are already gauge invariant), while the
mass and Higgs coupling terms give rise to a kinetic term for the $\zeta_i$
(stemming from the gauge boson mass term) and to
nonpolynomial interactions of the
St\"uckelberg scalars to the physical particles. Thereby these terms
have become replaced by gauge invariant expressions.

Two important points should be mentioned here.
\begin{itemize}
\item
Using the St\"uckelberg formalism one can construct gauge theories with massive
vector bosons. Such theories are usually described by using the formalism of
spontanous symmetry breaking (SSB).
Note that the St\"uckelberg formalism does not avoid SSB, since the
unitary matrix $U$ has nonvanishing VEV,
\be UU^\dagger={\bf 1}\quad\Rightarrow\quad
\langle U\rangle_0={\bf 1}. \label{constraint}\ee
But here, the nonvanishing VEV is not realized by introducing a
scalar-self-interaction potential with a nontrivial minimum
but by  imposing \eref{constraint} as a constraint, as in the gauged
nonlinear $\sigma$-model \cite{bash,apbe,lon}.
\item The starting U-gauge Lagrangian seems to be nonrenormalizable by simple
power counting arguments
(although we know it is renormalizable) due to the bad high energy
behaviour of the massive-vector-boson propagators. The St\"uckelberg formalism
introduces gauge freedom and enables so to perform calculations in the
\rx -gauge where the vector boson propagator behaves well. However this
formalism introduces
nonpolynomial interactions of the St\"uckelberg scalars $\zeta_i$.
I.~e., new interactions, arising from the expansion of $U$ \eref{u} in powers
of $\zeta_i$, have to be considered at each loop order, which again makes the
model nonrenormalizable by naive power counting.
So the problem of (possible) bad behaviour of loop corrections is only shifted
from the physical to the unphysical sector of the theory.
\end{itemize}

In the original St\"uckelberg formalism this is the end of the story, since
point transformations like \eref{stue} are not applied to the U-gauge of a
SBGT but to a Yang--Mills theory with mass terms added by hand. The
St\"uckelberg formalism transforms such a model to a
gauged nonlinear $\sigma$-model, which is
in fact nonrenormalizable \cite{apbe,lon,shiz},
since it is not possible to
reparametrize the unphycal scalars \eref{u} in a linear form to avoid
nonpolynomial interactions \cite{burn}.
However, in our case the starting Lagrangian contains
an extra Higgs boson (or several Higgs bosons in the case of more
extended theories). It has been introduced to ensure good high energy
behaviour on tree level and, in fact, it makes the model renormalizable, too,
since {\em physical and unphysical scalars together\/}
can be rewritten as a linear
expression by means of the point transformation \eref{nonlin}.
So in the linear parametrization (\eref{higgslin} with \eref{higgsshift})
 the nonpolynomial interactions
have been removed.
It has been
shown that for each tree unitary theory such a transformation exists
\cite{clt}.
The Jacobian determinant of \eref{nonlin} removes the extra term \eref{extra}.

So finally we have ``recovered'' the SM from its unitary gauge by applying
appropriate point transformations to the physical fields.
On the level of the classical Lagrangian, this has originally
been done in \cite{clt}. In  a quantum field theoretical treatment
performed here one has to consider the integration
measure of the PI, too, so that two new features arise:
\begin{itemize}
\item Field enlarging is easily achieved by mutiplying an infinite constant
(as \eref{fieldenlarge}) to the PI.
\item The starting Lagrangian has neccessarily to contain the extra term
\eref{extra} to cancel against the Jacobian determinant which arises as a
consequence of the the transformation of the functional integration measure.
\end{itemize}
As the result of this section we see, that a Higgs model can be derived from
the requirement of good high energy behaviour of tree-level
amplitudes and of loops without explicitly using the Higgs mechanism (although
SSB is implicitly included in the derivation). Thus, the physical sector of a
SBGT ``contains'' the entire model. Non-Abelian St\"uckelberg models are
renormalizable if, in addition to the unphysical St\"uckelberg
scalars, also physical scalars with appropriate couplings to the other
particles and themselves are present.


\section[]{Effective-Lagrangian Theories}
As mentioned before, the original purpose of the St\"uckelberg formalism was to
construct (gauge)
theories with massive gauge bosons but without physical scalars
\cite{stue,kugo,sots,dtt,burn}.
To construct a (non-Abelian) St\"uckelberg model one
starts from a Yang-Mills theory, introduces a mass term by hand, thereby
breaking gauge invarince explicitely, and then performes a field enlarging
St\"uckelberg transformation as \eref{stue} (after introducing the functional
integration over the unphysical St\"uckelberg scalars into the PI)
to restore gauge invariance.
Due to the Yang--Mills symmetry of the Lagrangian, except for the mass terms,
the St\"uckelberg transformation, which formally looks like a gauge
transformation with the gauge parameters being replaced by the unphysical
scalar fields,
affects only the mass terms in a nontrivial way thereby yielding, besides a
kinetic term for the St\"uckelberg scalars, nonpolynomial
interactions of the unphysical with the physical fields. In the Yang--Mills
part of the Lagrangian the original physical fields simply are replaced by the
transformed fields. Thus, the addition of unphysical scalars with suitable
nonpolynomial couplings (in the non-Abelian case)
to the physical fields embeds the original massive
Yang--Mills theory into an equivalent gauge theory. The resulting model
is a gauged nonlinear $\sigma$-model. As it is well known, this  model
can be derived in three
alternative ways (described here for the case of the \suu\ symmetry):
\begin{itemize}
\item By applying a St\"uckelberg transformation \eref{stue}
to the massive Yang--Mills theory
\cite{kugo} as explained above.
\item By imposing SSB in form of a constraint \eref{constraint}
on the scalar fields, which implies the parametrization \eref{u},
and gauging the broken symmetry \eref{zetatrafo}
by coupling $U$ minimally to the gauge fields \cite{bash}.
\item The gauged nonlinear
$\sigma$-model also turns out to be the limit of the SM for infinite
Higgs mass \cite{apbe,lon}. To perform this limit,
one has to substitute in the SM Lagrangian \eref{linv}
\be \Phi\to \frac{v}{\sqrt{2}}U\label{sigmatrafo}\ee
(with $\Phi$ and $U$ given by \eref{higgslin}, \eref{higgsshift} and \eref{u}),
which automatically removes the scalar self-interaction term
\be -V(\Phi)=-\frac{1}{2}\mu^2\tr(\Phi\Phi^\dagger)-
\frac{1}{4}\lambda\tr(\Phi\Phi^\dagger)^2
\label{higgsself}\ee and imposes the SSB \eref{constraint}.
\end{itemize}
Although the gauged nonlinear $\sigma$-model
is nonrenormalizable due to the nonpolynomial interactions,
gauge freedom enables perturbative calculations in the \rx -gauge and so one
finds, that the loops do not diverge as severe as one would expect from naive
power counting. In fact, the one-loop divergences of the gauged nonlinear
$\sigma$-model are only logarithmically cut-off dependent \cite{apbe,lon}.

During the last years electroweak effective-Lagrangian
theories with massive vector bosons and
non-Yang--Mills interactions have been studied as alternative models to the SM
(for examples see
\cite{zerw} and references in \cite{bulo}). These are also not gauge invariant,
because the mass terms
and the anomalous interactions violate gauge invariance. However, recently it
has been found that each effective-Lagrangian
theory is equivalent to a gauge theory
\cite{alda} , i.~e., it can be written as a gauge theory by performing a field
enlarging point transformation.
For the case of theories containing electroweak vector bosons with
arbitrary self-interactions, this was investigated in \cite{bulo}. The
transformation which was used for reformulating such a theory as a SBGT can be
identified with the St\"uckelberg transformation \eref{stue}, as one can see
by the same reasoning as above. Thereby, the spontanous symmetry
breaking is realized nonlinearly.
Again, the effect of the non-gauge-invariant
terms is cancelled by appropriate couplings of unphysical scalars but,
in difference to a simple massive Yang--Mills theory, not only the mass terms
but also the anomalous interaction terms
give rise to new nonpolynomial interactions. So one can perform loop
calculations within models containing arbitrary vector-boson
self-interactions in the \rx -gauge and has better opportunities to subdue
the divergences in such a nonrenormalizable model.
The U-gauge of this SBGT becomes the original effective Lagrangian.

A method of deriving anomalous self-couplings of electroweak
vector bosons from a gauge
invariant Lagrangian with {\em linearly} realized symmetry (Higgs model)
is to add to the the SM Lagrangian extra dimension six (or higher) \suu\
invariant interaction terms, which contain the
anomalous couplings \cite{llr,buwy,ruj,haze,heve,gore,gkks}.
Although there are no nonpolynomial
interactions, these models are nonrenormalizable, too, due to the higher
dimension of the extra interaction terms. But it has
recently been shown (for special cases with extra dimension six terms)
that loop corrections depend (after renormalization)
only logarithmically on the cut off due to the linear realized gauge invariance
\cite{ruj,haze,heve}.
Although these terms lead to observable deviations from the SM
predictions (stemming from loop contributions to presently measurable
quantities), the empirical limits on such
deviations do not severely restrict the size of the
anomalous couplings deriving from the extra terms
because of the smallness of the loop contribution due to the
weak cut-off dependence \cite{ruj,haze,heve}.
In fact, most of the additional terms contain extra
interactions of the Higgs boson to the gauge bosons and the Higgs boson
contribution cancels the quadratic loop divergences \cite{haze}.
So we suppose that the St\"uckelberg
models \cite{bulo} will yield larger loop corrections and thus be suppressed
more strongly by present experiments since there is no Higgs scalar.

However, the two applied methods to
get anomalous interactions out of gauge invariant models are in principle
different. In \cite{llr,buwy,ruj,haze,heve,gore,gkks} terms
are added to the SM Lagrangian, most of them contain not only vector-boson
self-interaction terms but also couplings to the physical Higgs bosons.
So the original
effective Lagrangian is {\em extended\/} to a gauge invariant model
which is not equivalent to the original one, since there are additional Higgs
couplings.
In contrary, the generalized St\"uckelberg formalism  \cite{bulo},
i.~e., the introduction of nonlinearly realized symmetry enables to
express an effective Lagrangian in terms of an {\em equivalent\/} gauge theory.
{}From the above discussion it is clear how these two formalisms are
connected. Given an effective Lagrangian for electroweak
theory with arbitrary vector-boson self-interactions,
one first performs the St\"uckelberg
transformation \eref{stue} and finds the equivalent generalized gauged
nonlinear $\sigma$-model. This can, like the usual gauged nonlinear
$\sigma$-model, be understood as the limit $M_H\to\infty$ of a Higgs model
with linearly realized symmetry \cite{apbe,lon}. To recover the Higgs
model one has to replace, reversing \eref{sigmatrafo},
\be U\to\frac{\sqrt{2}}{v}\Phi\label{recov}\ee
(with \eref{higgslin}, \eref{higgsshift})\footnote{It should be clear that
\eref{recov} cannot be understood  as a field enlarging point transformation
like \eref{stue}, since a hermitian matrix
$\Phi$ \eref{higgslin} can in general not be
expressed in terms of a unitary matrix $U$ \eref{u}. This shows that the step
from the generalized gauged nonlinear $\sigma$-model to the generalized
Higgs model is indeed an extension of the theory.}
and to add the Higgs potential \eref{higgsself},
which implies a nonvanishing VEV and replaces the constraint
\eref{constraint}. As in the previous section the addition of a physical Higgs
boson enables a linear parametrization of the scalar sector and removes the
nonpolynomial interactions.
By this simple formalism one can extend {\em each} effective-Lagrangian
electroweak theory with arbitrary self-interactions
to an
\suu\ invariant theory with linearly realized symmetry, which is expected
to yield a more decent loop behaviour.
This has explicitly been constructed for the special case of arbitrary cubic
self-interactions in \cite{gore}. Here, we have given the general formalism to
understand and perform this embedding for all types of vector boson
self-interactions\footnote{The investigation of this section
concerning arbitrary vector-boson self-interactions also can be
applied to get arbitrary fermionic interactions out of a (linearly or
nonlinearly realized) SBGT. However, we do not
stress this point here because this is of less phenomenological
interest since the SM fermionic interactions are very well confirmed in
experiments now.}.
Such a model will in general contain extra interaction
terms of even higher dimension than six \cite{gore}, which are, however,
supposed to yield  higher loop divergences and so to be more suppressed by
their indirect effects on present experiments than the dimension six terms.


\section[]{Summary}
In this paper we have reinvestigated
the three different methods to reduce a SBGT
to its physical sector (U-gauge Lagrangian)
within the formalism of Lagrangian PI. In difference to
the naive classical U-gauge,
the quantum U-gauge Lagrangian contains an extra nonpolynomial
quartic divergent self-interaction term of the Higgs boson(s) \eref{extra},
which is
necessary to make the theory manifestly
renormalizable even in the U-gauge. This term can be interpreted
as a remnant of the ghost term if the U-gauge is constructed by gauge
fixing
or by \rx -limiting procedure or it derives
from the functional Jacobian
determinant if the U-gauge is constructed by St\"uckelberg transformations.
Fortunately, for the phenomenologically most interesting processes
the extra term \eref{extra} is irrelevant in one-loop calculations.

On the basis of this analysis
we have shown how to construct SBGTs from effective-Lagrangian
models such that
the original Lagrangian turns out to be the U-gauge. In fact, each
effective-Lagrangian theory is equivalent to a SBGT with {\em nonlinearly\/}
realized
symmetry and each effective-Lagrangian theory, which is tree unitary, thus
containing (a) physical Higgs boson(s), and in which all quartic divergent loop
implied Higgs self couplings are removed by extra counterterms,
is equivalent to a SBGT with {\em linearly\/}
realized symmetry broken by the Higgs mechanism. Therefore,
gauge freedom is nothing
special for massive-vector-boson theories. This has
important consequences for the present phenomenolocial discussion.

Within this context the old-fashioned St\"uckelberg formalism has acquired new
importance since it can be implemented into the modern formalism of
gauge theories: on the one hand
it can be used to construct effective gauge theories
with nonlinearly realized symmetry,
containing no physical Higgs
bosons, which
are extentions of the gauged
nonlinear $\sigma$-model with anomalous gauge-boson self-intercations; on the
other hand, the existence of physical Higgs bosons
with appropriate coupling structure
implied by the demand of tree unitarity (and of a suitible quartic divergent
nonpolynomial Higgs self-interaction term) enables to find a linear
representation of the scalar fields implying a renormalizable
St\"uckelberg model.
Introducing Higgs bosons by hand, the St\"uckelberg formalism can even be used
to derive arbitrary vector-boson self-interactions from Higgs models
with linearly realized symmetry.

For treating
the point transformations of the fields in the present paper we always
had to consider the Jacobian determinants of the functional integration
measure. A more natural way would be to treat this subject in Hamiltonian
instead of Lagrangian PI formalism \cite{fadd} since point tranformations are
canonical transformations in phase space that do not change the functional
intergration measure if integration is performed over fields and conjugate
fields. This will be the subject of a forthcoming paper.


\section*{Acknowledgement}
One of us (C. G.-K.) thanks D. Schildknecht for arising his interest in the
unitary gauge and the St\"uckelberg formalism.
We thank J. S{\l}adkowski and G. J. Gounaris for helpful discussions.




\section*{Figure Captions}
\newcounter{fig}
\newlength{\fig}
\newlength{\numnum}
\settowidth{\numnum}{\bf 1}
\settowidth{\fig}{\bf Figure 1:}
\begin{list}{\bf Figure \makebox[\numnum][r]{\arabic{fig}}:}{\usecounter{fig}
\labelwidth\fig \leftmargin\labelwidth \addtolength\leftmargin\labelsep}
\item \label{feynrulesug}Feynman rules obtained from the ghost term \eref{ufp}
in the U-gauge. In all figures the solid lines represent the Higgs lines and
the dotted ones the ghost lines.
\item \label{ghostloop}Ghost loop connected to $N$
Higgs lines contributing to the
Feynman diagrams in the U-gauge. The internal ghosts may be $\eta^\pm$ or
$\eta_Z$.
\item \label{extravertex}Extra quartic divergent $N$-Higgs boson vertex.
\item \label{feynrx}Feynman rules for Fig.~\ref{ghostloop} in the \rx -gauge.
\end{list}

\begin{thebibliography}{99}
\bibitem{feyn}R. P. Feynman, Rev.\ Mod.\ Phys.\ {\bf 20} (1948) 367
\bibitem{fapo}L. D. Faddeev and V. N. Popov, \phlb{25} (1967) 29
\bibitem{higgs}P. W. Higgs, \phr{145} (1966) 1156
\bibitem{kibb}T. W. B. Kibble, \phr{155} (1967) 1554
\bibitem{lezj}B. W. Lee and J. Zinn-Justin, \phrd{5} (1972) 3121, 3137, 3155,
{\bf D7} (1973) 1049
\bibitem{fls}K. Fujikawa, B. W. Lee and A. I. Sanda, \phrd{6} (1972) 2923
\bibitem{able}E. S. Abers and B. W. Lee, \phrp{9} (1973) 1
\bibitem{wein2}S. Weinberg, \phrd{7} (1973) 2887
\bibitem{leya}T. D. Lee and C. N. Yang, \phr{128} (1962) 885
\bibitem{sast}A. Salam and J. Strathdee, \phrd{2} (1970) 2869
\bibitem{wein1}S. Weinberg, \phrd{7} (1973) 1068
\bibitem{jogl}S. D. Joglekar, \aph{83} (1974) 427
\bibitem{clt}J. M. Cornwall, D. N. Levin and G. Tiktopoulos,
\phrd{10} (1974) 1145
\bibitem{stue}E. C. G. St\"uckelberg, Helv. Phys. Acta {\bf 11} (1938) 299
\bibitem{kugo}T. Kunimasa and T. Goto, \ptph{37} (1967) 452
\bibitem{sots}T. Sonoda and S. Y. Tsai, \ptph{71} (1984) 878
\bibitem{dtt}R. Delburgo, S. Twisk and G. Thompson, Int.\ Journ.\ of Mod.\
Phys.\ {\bf A3} (1988) 435
\bibitem{llsm}C. H. Llewellyn Smith, \phlb{46} (1973) 233
\bibitem{alda}J. Alfaro and P. H. Damgaard, \aph{202} (1990) 398
\bibitem{bulo}C. P. Burgess and D. London, McGill-Preprints McGill-92/04,
McGill-92/14 (1992)
\bibitem{sm}S. Weinberg, \phrl{19} (1967) 1264
\bibitem{sm2} A. Salam, Proc.\ 8th Nobel Symposium, ed.\ N. Svartholm (Almquist
and Wiksells, Stockholm, 1968) p.\ 367
\bibitem{balo}D. Bailin and A. Love, ``Introduction to Gauge Field Theory'',
Hilger (1986), Chapter 14
\bibitem{bash}W. A. Bardeen and K. Shizuya, \phrd{18} (1978) 1969
\bibitem{apbe}T. Appelquist and C. Bernard, \phrd{22} (1980) 200
\bibitem{lon}A. C. Longhitano, \nphb{188} (1981) 118
\bibitem{shiz}K. Shizuya, \nphb{121} (1977) 125
\bibitem{burn}A. Burnel, \phrd{33} (1986) 2985
\bibitem{zerw}``$\rm e^+e^-$ Collisions at 500 GeV, The Physics Potential'',
ed.\ P. W. Zerwas, DESY-Preprint DESY 92-123 (1992)
\bibitem{llr}C. N. Leung, S. T. Love and S. Rao, \zphc{31} (1986) 433
\bibitem{buwy} W. Buchm\"uller and D. Wyler, \nphb{268} (1986) 621
\bibitem{ruj}A. de R\'{u}jula, M. B. Gavela, P. Hern\'{a}ndez and E. Mass\'{o},
\nphb{384} (1992) 3
\bibitem{haze}K. Hagiwara, S. Ishihara, R. Szalapski and D. Zeppenfeld,
\phlb{283} (1992) 353
\bibitem{heve}P. Hern\'{a}ndez and F. J. Vegas, CERN-Preprint CERN-TH 6670
(1992)
\bibitem{gore}G. J. Gounaris and F. M. Renard, Montpellier-Preprint
PM/92-31 (1992)
\bibitem{gkks}C. Grosse-Knetter, I. Kuss and D. Schildknecht, in preparation
\bibitem{fadd}L. D. Faddeev, Theor.\ Math.\ Phys.\ {\bf 1} (1969) 1
\end{thebibliography}
\end{document}